\documentclass[twocolumn,showpacs,preprintnumbers,amsmath,amssymb,prl,superscriptaddress]{revtex4}

\usepackage{amsmath,amssymb,amsthm,color}
\usepackage{graphicx}
\usepackage{dcolumn}

\newcommand {\beq}{\begin{eqnarray}}
\newcommand {\eeq}{\end{eqnarray}}

\begin{document}


\title{Gravity Dual for Cyclic Renormalization Group Flow without Conformal Invariance }

\author{Yu Nakayama}

\affiliation{California Institute of Technology, 452-48, Pasadena, California 91125, USA}


\begin{abstract}
We construct a gravity dual for scale invariant but non-conformal field theories with a cyclic renormalization group flow. A slight modification of our construction gives a gravity dual of discretely scale invariant field theories. The underlying gravitational theory breaks the null energy condition.

\end{abstract}

\maketitle

\section{1. Introduction}
Once there was a myth: scale invariance is equivalent to conformal invariance in unitary relativistic quantum field theories. It is true in (1+1) dimension under additional technical assumptions as proved in \cite{Zamolodchikov:1986gt}\cite{Polchinski:1987dy}. In higher dimensions, we had cherished the myth until quite recently since nobody had presented any convincing counterexamples. With the lack of counterexamples, they have tried to give a proof in higher dimension, and have gained some insights \cite{Dorigoni:2009ra}\cite{Antoniadis:2011gn}\cite{Zheng:2011bp}, but the proof has never yet come.

It was shown, however, in \cite{Jackiw:2011vz}\cite{ElShowk:2011gz} that free Maxwell field theory in $d\ge 5$ is scale invariant but not conformally invariant. This is the first example of scale invariant but non-conformal fixed point of the renormalization group in unitary relativistic field theories, and it was the first indication of the end of the myth.
 In a more recent paper \cite{Fortin:2011ks}, it was suggested that in $(4-\epsilon)$ dimension, scalar field theories coupled with fermions may show scale invariance but non-conformal invariance at two-loop order. The latter example is qualitatively different from the former not only because it is interacting but the theory claims to show a cyclic behavior along the renormalization group flow. In their example, the naive scale transformation must be accompanied with the field space rotation to construct the conserved scale transformation current.

The cyclic behavior of the renormalization group flow is of theoretical interest. The possibility was suggested in earlier literatures of the renormalization group (see e.g. \cite{Wilson:1973jj}), but a concrete example had never been found in relativistic field theories. The belief in the non-existence of such a behavior is tightly related with the ``c-theorem" \cite{Zamolodchikov:1986gt}: there exists a monotonically decreasing function along the renormalization group flow, which remains another myth in relativistic quantum field theories in higher dimensions \cite{Cardy:1988cwa}. The proposed example \cite{Fortin:2011ks}, if true, seems to open a novel possibility of the renormalization group flow, and would completely change our understanding of the renormalization group. Thus, it would be extremely important to find more concrete and conclusive examples. Searching for a gravity dual description via holography seems to be one approach we can take, which is complementary to the perturbative field theory analysis. The goal of this paper is provide such a gravity dual example. 

The possibility to construct a gravity dual for scale invariant but non-conformal field theories was pursued in \cite{Nakayama:2009qu}\cite{Nakayama:2009fe}\cite{Nakayama:2010wx} (see \cite{Nakayama:2010zz} for a review). The conclusion is it is a no-go within the conventional Einstein gravity (with possible higher derivative corrections) unless we violate the strict version of null energy condition. The scale invariance dictates that the metric must be AdS, and a possible violation in the matter sector then violates the energy condition.

As a consequence, in our search for the gravity dual, we dare to abandon the null energy condition. At the sacrifice of the energy condition, we will succeed in constructing a gravity dual for scale invariant but non-conformal field theories with a cyclic renormalization group flow. A qualitative feature of our model may be compared with the $(4-\epsilon)$ dimensional field theory construction of \cite{Fortin:2011ks}.

It turns out that a slight modification of our construction also gives a gravity dual for discretely scale invariant field theories. The dual field theory is invariant under the scale transformation $ x_i \to e^{\lambda} x_i$, but with a discrete set of numbers $\lambda$ (obviously $\lambda$ here forms a discrete Abelian group $\mathbb{Z}$). As far as we know, there have been no relativistic field theories that show such a behavior (we note that in the non-relativistic setup, the existence of a cyclic renormalization group flow is established and the gravity dual was studied e.g. in \cite{Kaplan:2009kr}\cite{Moroz:2009kv}\cite{Wen:2010et}).  Our gravity construction suggests that such a theory may be realizable, and it should share the same status of generic scale invariant but non-conformal field theories in the landscape of relativistic field theories, like the site of Troy whose existence had been dubious but turned out to exist.

The organization of the paper is as follows. In section 2, we extend our previous model of a gravity dual for scale invariant but non-conformal field theories, and investigate more generic solutions. We show that these solutions correspond to scale invariant but non-conformal field theories with a cyclic renormalization group flow. We further show that a slight modification of our construction gives  a gravity dual of discretely scale invariant field theories. In section 3, we discuss several physical aspects of our gravity description, and suggest possible future studies.

\section{2. Construction of Gravity Dual}

Our construction of a gravity dual for scale invariant but non-conformal field theories with a cyclic renormalization group flow is based on the generalization of the model studied in \cite{Nakayama:2009qu}. The salient feature of the model that we will exploit here is that a certain non-trivial profile of the matter does not source the energy-momentum tensor. Such a property is inconsistent with the strict null energy condition, but as shown in \cite{Nakayama:2009fe}\cite{Nakayama:2010wx}, the violation is necessary to construct the scale invariant but non-conformal field theories within the conventional Einstein gravity. 

\subsection{2.1 Model}

The model is based on the $d+1$ dimensional Einstein gravity with a negative  cosmological constant $\Lambda$ coupled with a complex valued 1-form field $A = A_\mu dx^\mu$. The action is given by
\begin{align}
\int d^{d+1}x \sqrt{g} \left(\frac{1}{2}R - \Lambda-\frac{1}{2}|F_{\mu\nu}|^2 -\sum_n \frac{g_n}{n} |A_\mu|^{2n} \right) \ , \label{action}
\end{align}
where  we have chosen the Maxwell-like kinetic term for the complex valued 1-form field with $F_{\mu\nu} = \partial_\mu A_\nu - \partial_\nu A_\mu$.

The vacuum solution of the Einstein equation is given by the AdS space ($i = 1, \cdots d$)
\begin{align}
ds^2 = \frac{dz^2 + dx_i dx^i}{z^2} \ .
\end{align}
We arrange the coefficients in the potential $g_n$ so that 
\begin{align}
 A = \frac{a dz}{z} \label{orig1}
\end{align}
with non-zero $a$ is a solution of the equation of motion. It is easy to see that the field configuration \eqref{orig1} does not source the energy-momentum tensor and the AdS space is still a solution of the Einstein equation. 

The field configuration \eqref{orig1} is invariant under the scale transformation $(z, x_i) \to (e^{\lambda} z, e^{\lambda} x_i)$ with real $\lambda$, but is not invariant under the isometry transformation ($\delta z = 2(v^j x_j)z, \delta x^i = 2(v^j x_j) x^i - (z^2 + x^j x_j)v^i$ ) corresponding to the special conformal transformation with the infinitesimal parameter $v_i$.   In this way, this model with the particular solution \eqref{orig1} represents a gravity dual for scale invariant but non-conformal field theories \cite{Nakayama:2009qu}.

Our model admits more general solutions than \eqref{orig1}. Indeed
\begin{align}
 A = \frac{a dz}{z} z^{i \theta} \label{orig}
\end{align}
for any real $\theta$ solves the equation of motions in the AdS space. The solution is not invariant under the geometric scale transformation  $(z, x_i) \to (e^{\lambda} z, e^{\lambda} x_i)$, but it transforms the 1-form field as $A \to A e^{i\theta \lambda}$. In the following discussion, the phase shift under the scaling transformation plays a significant role. On the other hand, under the special conformal transformation, the 1-form field $A$ does not transform nicely, so the dual field theory must break the special conformal invariance.

\subsection{2.2 Geometry and field theory interpretations}
As we have seen, the general solutions \eqref{orig} are not invariant under the simple geometric scaling transformation, but we would like to argue that we can augment the internal symmetry transformation to cancel the phase to recover the scale invariance. The situation depends on the internal symmetry in three cases:

(A) The internal symmetry is global. Suppose that the action has a global phase rotation symmetry for the 1-form field $A \to e^{i\phi} A$. The simple action \eqref{action} indeed possesses this symmetry. Then, we can undone the phase shift $A \to e^{i\theta \lambda} A$ associated with the scale transformation $(z, x_i) \to (e^{\lambda} z, e^{\lambda} x_i)$ by the global phase rotation. Thus, the model is invariant under the combined symmetry of the geometric scale transformation and the global phase rotation. Note that under the global symmetry transformation, non-trivial states are transformed to equivalent states, but the new states are not identical to the original ones.
Therefore, the scaling transformation gives a translation along the renormalization group trajectory, and the renormalization group trajectory here is cyclic, given by the phase rotation in the field space. 
This precisely corresponds to the scale invariant but non-conformal cyclic renormalization group trajectory proposed in \cite{Fortin:2011ks}.

Our model has a parameter $\theta$ which is not fixed. Thus, our model shows a class of cyclic renormalization group trajectories parametrized by $\theta$. For a given $\theta$, the physics is the same along the cyclic renormalization group trajectory as the scale invariance suggests, but for each $\theta$, the theory and the renormalization group flow possesses different physical properties: $\theta$ determines how much the field rotation is mixed with the scale transformation. In other words, the parameter $\theta$ is an exactly marginal deformation of the dual scale invariant but non-conformal field theories.
In this sense, our model has a two-dimensional fixed ``plane" under the renormalization group rather than the fixed ``loop" in the model of \cite{Fortin:2011ks}. We note, however, that we may always modify and tune the higher derivative terms in the kinetic term of the action \eqref{action} so that only a specific value of $\theta$ solves the equation of motion. As a result, the existence of the extra parameter $\theta$ is not always necessary in the gravity construction.

We may want to compare the qualitative features of our model with the one studied in \cite{Fortin:2011ks} from the perturbative field theory perspective. For instance, the computation of correlation functions via GKPW prescription does not give a simple scaling correlation functions for the operator dual to $A_\mu$, but we have to rotate the basis of the operator along the renormalization group flow to make it manifestly scale invariant. This is completely in agreement with the cyclic behavior of the renormalization group flow studied in \cite{Fortin:2011ks}.

Our cyclic renormalization group trajectory does not violate the gravitational analogue of c-theorem \cite{Girardello:1998pd}\cite{Freedman:1999gp} but only marginally: the central charge is not monotonically decreasing but rather it stays constant. Since the effective vacuum energy retains the same value along the cyclic renormalization group flow along the $z$ direction, the gravitational central charge remains the same. This is expected because the scale invariance dictates that the two-point function of the energy-momentum tensor must scale and its coefficient must be independent of the scale. We, nevertheless, point out that our model violates the null energy condition that is relevant for the derivation of the gravitational c-theorem. Therefore, it is not obvious whether the gravitational c-theorem still holds once we stay away from the scale invariant renormalization group trajectory.

(B)  The internal symmetry is local. When the phase rotation symmetry of the 1-form field is a gauge symmetry, the discussion becomes slightly different from the global case discussed in (A).  The local gauge invariance introduces the additional vector potential $V = V_\mu dx^\mu$ which transforms as $ V \to V + d \alpha$  under $A \to e^{i\alpha} A$.
 To understand the scale invariance, we can always undone the phase shift $A \to e^{i\theta \lambda} A$ associated with the geometrical scale transformation $(z, x_i) \to (e^{\lambda} z, e^{\lambda} x_i)$ by the gauge transformation $A \to e^{i\alpha} A$. Unlike the global symmetry case discussed in (A), however, the scale transformation gives the physically identical states to the original ones due to the gauge identification, so the renormalization group flow has a fixed point for every $\theta$, and does not show the cyclic behavior. Thus, we have a gravity dual description of scale invariant but non-conformal field theories sitting at the scale invariant fixed point.

Alternatively, we can fix the gauge so that \eqref{orig} becomes $A = a\frac{dz}{z}$, with a non-zero gauge field $V = -\theta\frac{dz}{z}$.  The field configuration is essentially same as the one studied in  \cite{Nakayama:2009qu}, and it is scale invariant, but not conformally invariant. With this picture, it is obvious that it does not show a cyclic behavior. The parameter $\theta$ plays  a role of the moduli, and a different $\theta$ describes a different scale invariant but non-conformal fixed point.

(C) The continuous internal symmetry does not exist. It is possible that via quantum gravity effects, the continuous symmetry may get broken to its discrete subgroup. Or, from the beginning there is no logical reason to impose the internal symmetry for the 1-form vector field as long as \eqref{orig} is a solution of the equation of motion.
A key observation is that even if the continuous symmetry did not exist, the field configuration \eqref{orig} would be invariant at least under the discrete scale transformation for $\theta\lambda \in 2 \pi \mathbb{Z}$. Thus, the dual field theories must be invariant (only) under the discrete scale transformation. Obviously, the conformal invariance is broken there. We note that there is no analogue of the Zamolodchikov-Polchinski theorem (and its higher dimensional conjecture) for the discrete scale invariance.

This opens up a new class of renormalization group structure (i.e. the discrete scale invariance), so it is worth studying some details of its implication from the dual gravity perspective. As a toy example, let us couple our model \eqref{action} with a scalar field $\phi$ via the interaction that breaks the continuous phase rotation symmetry
\begin{align}
\int d^{d+1} x \sqrt{g} \left( \partial^\mu \phi \partial_\mu \phi + m^2 \phi^2 + \epsilon (A^\mu A_\mu + A^{*\mu} A^*_{\mu})\phi^3 \right) \ .
\end{align}
Note that at $\phi = 0$, the field configuration \eqref{orig} still solves the equations of motion, but the interaction now breaks the continuous scale symmetry down to its discrete version $\theta\lambda \in \pi \mathbb{Z}$, so the dual field theory must be only invariant under the discrete scale transformation.

We would like to study the correlation functions of the boundary operator $\mathcal{O}(x)$ which is dual to the scalar $\phi$ by using the standard GKPW prescription. It must show discrete scale invariance in their behavior.
We will demonstrate it for the three-point function. It can be computed order by order in $\epsilon$.
 The first order term (up to an overall normalization constant which we will not care) is given by
\begin{align}
&\langle \mathcal{O}(x_1) \mathcal{O}(x_2) \mathcal{O}(x_3)  \rangle \sim  \epsilon a^2 \int \frac{d^dx  dz}{z^{d+1}} \cos(2\theta\log z) \cr 
& \left(\frac{z}{(x- \vec{x}_1)^2} \right)^{\Delta} \left(\frac{z}{(x-\vec{x}_2)^2} \right)^{\Delta} \left(\frac{z}{(x-\vec{x}_3)^2} \right)^{\Delta}  \ , \label{correction}
\end{align}
where $\Delta = \frac{d}{2} + \frac{1}{2}\sqrt{d^2 + 4m^2}$, and $(x-\vec{x}_a)^2 = z^2 + (\vec{x}- \vec{x}_a)^2$.

Indeed, we can check that the expression scales under the discrete scale transformation $x_i \to e^{\lambda} x_i$, only when $ \theta \lambda \in \pi \mathbb{Z}$ (as $\sim e^{-3\lambda \Delta}$), and the continuous scaling invariance (as well as the conformal invariance) of the three-point function is explicitly broken in this expression. In general, we expect that the three-point functions of the discretely scale invariant field theories are given by
\begin{align}
&\langle \mathcal{O}(x_1) \mathcal{O}(x_2) \mathcal{O}(x_3) \rangle = \cr 
& \sum_{\Delta_1 + \Delta_2 + \Delta_3 = 3\Delta} \frac{f_{\Delta_1\Delta_2\Delta_3}(x_1-x_2, x_2-x_3, x_3-x_1)}{(x_1 - x_2)^{\Delta_1} (x_2 - x_3)^{\Delta_2} (x_3 - x_1)^{\Delta_3} }   \ , \label{core}
\end{align}
with a discretely scale invariant function $f_{\Delta_1\Delta_2\Delta_3}(e^{\lambda} x_1, e^{\lambda} x_2, e^{\lambda} x_3) = f_{\Delta_1\Delta_2\Delta_3}(x_1,x_2,x_3)$ for $ \theta \lambda \in \pi \mathbb{Z}$ with symmetric indices $\Delta_i$, which is consistent with the gravity dual behavior \eqref{correction}.
We conclude that our model is able to provide a gravity dual of discretely scale invariant field theories.

Note that the discrete scale invariance does not require the existence of any conserved current, so the condition for such a symmetry is much weaker than that for the continuous scale invariance or conformal invariance. Thus, naively we might expect that such examples are ubiquitous, although in reality they seem very scarce. We have not been aware of any argument that the discrete scale invariance is impossible in field theories (for instance, the behavior of the three-point function \eqref{core} is completely consistent with the axioms of unitary relativistic field theories unless $f(x)$ is too wild) except for possible conflict with the c-theorem, and we believe that this gravity example can be regarded as the first existence proof.

\section{3. Discussion and Conclusion}

In this paper, we have studied a gravitational realization of scale invariant but non-conformal field theories with a cyclic renormalization group flow. We also have shown that a slight modification of the model gives a gravity dual of  discretely scale invariant field theories. All these field theories sound exotic but we do not know any convincing argument against them. 

In order to realize scale invariant but non-conformal field theories from the holography, we have to break the strict version of the null energy condition. We do not know whether the violation of the null energy condition leads to pathology of our model, in particular upon quantization. Our starting action \eqref{action} is related to the (gauged) ghost condensation (see \cite{Nakayama:2009qu}), and it would be important to see how it is really consistent as a theory of quantum gravity, and whether it is possible to realize them in string theory. In the Euclidean signature, at the sacrifice of the reflection positivity, we can establish the background that shows  mixing between the geometric scale transformation and the internal rotation within the string theory \cite{Rey}. Note that even if the violation of the null energy condition were pathological in the Lorentzian signature, our model would be perfectly fine in the Euclidean signature.

One way to avoid the no-go theorem is to use non-Einstein type gravitational theory. An interesting example is higher spin gauge theory. For our purpose, we note that a realization of the gravity dual of a free scalar field theory was proposed in terms of the higher spin gauge theory in \cite{Douglas:2010rc}. A simple generalization is possible by considering free Maxwell theory instead of the free scalar. As shown in \cite{Jackiw:2011vz}\cite{ElShowk:2011gz}, the theory is scale invariant but not conformally invariant when $d\ge 5$. Clearly, the corresponding dual higher spin gauge theory is only invariant under scale isometry but not invariant under the special conformal transformation. This can be explicitly checked at the level of equations of motion. It is, however, difficult to understand the underlying covariant theory because the obtained equations of motion are based on a particular ansatz in a particular gauge. It would be interesting to pursue this problem further.

Finally, it is interesting to look for supersymmetric extensions. Indeed, Nahm's classification \cite{Nahm:1977tg} dictates that there is no superconformal algebra in $d \ge 7$ (unless we circumvent the Coleman-Mandula theorem). Thus, non-trivial interacting supersymmetric scale invariant field theories in higher dimension than seven, if any, cannot be conformally invariant while preserving the supersymmetry. The construction of such a gravity dual would be challenging.

We hope that a further study of this subject will demystify the myth: scale invariance is equivalent to conformal invariance in unitary relativistic field theories.

\section*{Acknowledgments}
We greatly thank Chiu-Man Ho for suggesting the possibility of discrete scale invariance in quantum field theories. We also acknowledge Soo-Jong Rey for discussions. The work is supported by Sherman-Fairchild Senior Research Fellowship at California Institute of Technology.



\begin{thebibliography}{99}
\bibitem{Zamolodchikov:1986gt}
  A.~B.~Zamolodchikov,
  JETP Lett.\  {\bf 43} (1986) 730
  [Pisma Zh.\ Eksp.\ Teor.\ Fiz.\  {\bf 43} (1986) 565].
\bibitem{Polchinski:1987dy}
  J.~Polchinski,
  Nucl.\ Phys.\  B {\bf 303}, 226 (1988).

\bibitem{Dorigoni:2009ra}
  D.~Dorigoni, V.~S.~Rychkov,
  
  [arXiv:0910.1087 [hep-th]].

\bibitem{Antoniadis:2011gn}
  I.~Antoniadis and M.~Buican,
  Phys.\ Rev.\  D {\bf 83}, 105011 (2011)
  [arXiv:1102.2294 [hep-th]].

\bibitem{Zheng:2011bp}
  S.~Zheng and Y.~Yu,
  arXiv:1103.3948 [hep-th].


\bibitem{Jackiw:2011vz}
  R.~Jackiw and S.~Y.~Pi,
  J.\ Phys.\ A  {\bf 44}, 223001 (2011)
  [arXiv:1101.4886 [math-ph]].



\bibitem{ElShowk:2011gz}
  S.~El-Showk, Y.~Nakayama and S.~Rychkov,
  Nucl.\ Phys.\  B {\bf 848}, 578 (2011)
  [arXiv:1101.5385 [hep-th]].


\bibitem{Fortin:2011ks}
  J.~F.~Fortin, B.~Grinstein and A.~Stergiou,
  arXiv:1106.2540 [hep-th].

\bibitem{Wilson:1973jj}
  K.~G.~Wilson, J.~B.~Kogut,
  Phys.\ Rept.\  {\bf 12}, 75-200 (1974).

\bibitem{Cardy:1988cwa}
  J.~L.~Cardy,
  Phys.\ Lett.\  {\bf B215}, 749-752 (1988).






\bibitem{Nakayama:2009qu}
  Y.~Nakayama,
  JHEP {\bf 0911}, 061 (2009).
  [arXiv:0907.0227 [hep-th]].

\bibitem{Nakayama:2009fe}
  Y.~Nakayama,
  JHEP {\bf 1001}, 030 (2010).
  [arXiv:0909.4297 [hep-th]].

\bibitem{Nakayama:2010wx}
  Y.~Nakayama,
  [arXiv:1009.0491 [hep-th]].

\bibitem{Nakayama:2010zz}
  Y.~Nakayama,
  Int.\ J.\ Mod.\ Phys.\  {\bf A25}, 4849-4873 (2010).

\bibitem{Kaplan:2009kr}
  D.~B.~Kaplan, J.~-W.~Lee, D.~T.~Son, M.~A.~Stephanov,
  Phys.\ Rev.\  {\bf D80}, 125005 (2009).
  [arXiv:0905.4752 [hep-th]].


\bibitem{Moroz:2009kv}
  S.~Moroz,
  Phys.\ Rev.\  D {\bf 81}, 066002 (2010)
  [arXiv:0911.4060 [hep-th]].
\bibitem{Wen:2010et}
  W.~Y.~Wen,
  arXiv:1009.3952 [hep-th].


  
\bibitem{Girardello:1998pd}
  L.~Girardello, M.~Petrini, M.~Porrati and A.~Zaffaroni,
  JHEP {\bf 9812}, 022 (1998)
  [arXiv:hep-th/9810126].

\bibitem{Freedman:1999gp}
  D.~Z.~Freedman, S.~S.~Gubser, K.~Pilch, N.~P.~Warner,
 Adv.\ Theor.\ Math.\ Phys.\  {\bf 3}, 363-417 (1999).
  [hep-th/9904017].

\bibitem{Rey}
Y.~Nakayama and S.~Rey, work in progress

\bibitem{Douglas:2010rc}
  M.~R.~Douglas, L.~Mazzucato, S.~S.~Razamat,
  Phys.\ Rev.\  {\bf D83}, 071701 (2011).
  [arXiv:1011.4926 [hep-th]].


\bibitem{Nahm:1977tg}
  W.~Nahm,
  Nucl.\ Phys.\  {\bf B135}, 149 (1978).
  
\end{thebibliography}
\end{document}